\documentclass[twoside]{article}
\usepackage[latin1]{inputenc}
\usepackage{t1enc}
\usepackage{a4}
\usepackage{tabularx}
\usepackage{epsf}
\usepackage{psfig}

\def\bref{\vspace{4pt}\noindent\hangindent=10mm}

\begin{document}

\setcounter{figure}{0}
\setcounter{section}{0}
\setcounter{equation}{0}

\begin{center}
{\Large\bf
The FORS Deep Field}\\[0.2cm]
\bigskip

Jochen Heidt$^1$, Immo Appenzeller$^1$, Ralf Bender$^2$, Asmus B\"ohm$^3$,\\
Nive Drory$^2$, Klaus J. Fricke$^3$, Achim Gabasch$^2$, Ulrich Hopp$^2$,\\
Klaus J\"ager$^3$, Martin K\"ummel$^4$, D\"orte Mehlert$^1$,
Claus M\"ollenhoff$^1$,\\
 Alan Moorwoord$^5$, Harald Nicklas$^3$, Stefan Noll$^1$, Roberto Saglia$^2$,\\
 Walter Seifert$^1$, Stella Seitz$^2$, Otmar Stahl$^1$, Eckhard Sutorius$^1$,\\
Thomas Szeifert$^{1,6}$, Stefan J. Wagner$^1$, Bodo Ziegler$^3$\\[0.17cm]
$^1$ Landessternwarte Heidelberg, K\"onigstuhl, 69117 Heidelberg, Germany \\
$^2$ Universit\"atssternwarte M\"unchen, Scheinerstr. 1, 81679 M\"unchen,
Germany \\
$^3$ Universit\"atssternwarte G\"ottingen, Geismarlandstr. 11, 37083
G\"ottingen, Germany\\ 
$^4$ Max-Planck-Institut f\"ur Astronomie, K\"onigstuhl 17, 69117 Heidelberg,
Germany \\
$^5$ European Southern Observatory, Karl-Schwarzschild-Str. 2, 85748 Garching,
Germany\\
$^6$ European Southern Observatory Santiago, Alonso de Cordova 3107, Santiago
19, Chile\\
\end{center}

\vspace{0.5cm}

\begin{abstract}
\noindent{\it
Dedicating a major fraction of its guaranteed time, the FORS consortium 
established a FORS Deep Field which contains a known QSO at z = 3.36. 
It was imaged in UBgRIz with FORS at the VLT as well as in J and Ks 
with the NTT. Covering an area 6-8 times larger as the HDFs but with 
similar depth in the optical it is one of the largest  deep fields up to 
date to investigate i) galaxy evolution in the field from present up to 
z $\sim$ 5, ii) the galaxy distribution in the line of sight to the QSO, 
iii) the high-z QSO environment and iv) the galaxy-galaxy lensing 
signal in such a large field. In this presentation a status report of the FORS
Deep Field project is given. In particular, the field selection, 
the imaging results (number counts, photometric redshifts etc.) and 
the first spectroscopic results are presented.
}
\end{abstract}

\section{Introduction}

Deep fields are one of the most powerful tools to explore 
galaxy evolution over a wide redshift range. The main aim of  
this kind of studies is to constrain evolutionary scenario such as the 
hierarchical structure formation typical of Cold Dark Matter universes
(e.g. Kauffmann, 1996) or the picture of monolithic collapse, which 
assumes galaxy formation at very high redshifts and passive evolution
(e.g. Larson, 1974). Those studies are typically performed in several 
subsequent steps. Images with the best possible resolution, as deep and
covering an area as large as possible, are used to study the 
morphology of the objects of interest.
A combination of several optical and near-IR broad-band (and in some cases
narrow-band) images of a deep field allow to estimate photometric redshifts,
which in turn can be used to either preselect (classes of) objects
for follow-up spectroscopy or even constrain galaxy evolution without
spectroscopical confirmation for objects beyond the reach of current 
8 - 10 m class telescopes. Finally, spectroscopy of individual (classes of)
galaxies are used to derive their kinematics and to investigate
their chemical composition, star formation histories etc.

The Hubble Deep Field North (HDF-N, Williams et al., 1996) and related
follow-up observations e.g. with Keck 
dramatically improved our knowledge of
galaxy evolution in the redshift range z = 1 - 4. It is the deepest view
of the sky ever made, with excellent resolution, but it has only a relatively
small field of view ($\sim$ 5.6 sq.arcmin). In contrast, the Calar Alto
Deep Imaging Survey (CADIS, Meisenheimer et al., 1998) is less deep, but 
covers a much larger area (several 100 sq. arcmin) and searches specifically
for primeval galaxies in the redshift range z = 4.6 - 6.7. By now,
several other Deep Field studies have been conducted or are underway,
such as the NTT Deep Field (Arnouts et al., 1999), the WHT Deep Field
(McCracken et al. 2000) or the DEEP Survey (see e.g. Koo, 1999).
All these Deep Fields fall inbetween the HDF-N and CADIS in terms 
of deepness, resolution and area surveyed.

The FORS instruments for the VLT1 and VLT2 telescopes have been built
by a consortium of the State Observatory in Heidelberg, the
University Observatory in  G\"ottingen and the University Observatory 
in Munic (Appenzeller et al., 1992). For their efforts the three 
institutes received guaranteed observing 
time on the 2 VLT telescopes with the FORS instruments. In order to use the
guaranteed time efficiently, it was decided to spent a major fraction 
of the guaranteed time in establishing a FORS Deep Field (FDF). 
The advantages are obvious. The FDF conducted on a 8 m class telescope
allows imaging nearly as deep as is possible for the HDFs, with lower
resolution, but a 6 - 8 times larger field of view (providing better
statistics) and efficient follow-up multi-object spectroscopy 
with the same instruments. Moreover, it provides a large pool of data
for projects carried out within the three institutes involved,
which are supported financially by the government including two SFBs
(SFB 439 "Galaxies in the young universe in Heidelberg, 
SFB 375 "Astro particle physics" in Munic),
and the VW foundation programme ("Kinematic evolution of galaxies" in
G\"ottingen). 

In this article, a status report of the FDF project as of September
2000 is presented. 
In the following, the field selection will be described,
and the scientific aspects which will be addressed, will be summarized.
Then, the results from the imaging part will be presented,
followed by a short overview of the photometric redshifts. Finally, 
first spectroscopical results are shown and 
one of the scientific highlights so far briefly discussed.
The very first observations (selection of the FDF) started almost 
2 years ago, whereas the imaging part has been conducted in Fall 1999.
Since the annual meeting of the German Astronomical Society 2000 in Bremen,
the major part of the follow-up spectroscopy was also completed.
This subject, however, will only be touched briefly.

\section{Field selection}

A critical aspect for a project such as the FDF is the selection of a suitable
field. Owing to various constraints, the following selection criteria
were adopted:

\begin{itemize}

\item [1)] The field should be free of astronomical objects brighter than
19th mag (to avoid saturation during periods of excellent seeing conditions,
to avoid reflexes on the CCD and minimize overhead for readout, slewing etc.).

\item [2)] It should not contain any (known) galaxy cluster
(to avoid a distorsion of galaxy number counts).

\item [3)] The field should have good observability 
(close to zenith at Paranal).

\item [4)] In order to go as deep as possible and to allow observations in
  other wavebands low galactic extinction, low HI column density
and low IR-cirrus is required.

\item [5)] Strong x-ray sources and radio sources in the field should be 
avoided (to exclude potential galaxy clusters at high redshift).

\item [6)] No bright stars ($<$ 5mag) within $5^{o}$ should be present
(to avoid possible reflexes and straylight from the telescope structure).

\item [7)] A (radio-quiet) QSO at redshift $>$ 3 should be present in the 
field (to study QSO environment at high z and the IGM along the line of sight
to the QSO).

\end{itemize}

Due to the items 1, 4 and 6 the south galactic pole was a good region on 
the sky to start. 
Therefore all the QSO neighborhoods, which fulfilled criteria 3 and 
7 close to
the south galactic pole (and thus favouring criteria 1, 4 and 6) were selected
from the QSO catalogue of V\'{e}ron-Cetty
 \& V\'{e}ron (7. edition, 1997). This resulted in 32 candidates. 
Afterwards an  extensive search in
the literature and catalogs from radio up to x-ray bands, a visual check of
the DSS and photometry provided by the COSMOS scans was carried out, 
which revealed 4 promising candidates. These 4  candidates
were observed during the commissioning phase of FORS1, which showed that 
3 of them were not useful (they either contained
relatively bright galaxy clusters or had problems with the observability 
due to the absence of suitable guide stars at 
the VLT). Finally, a field with the
center coordinates $\alpha_{2000} = 01^{\rm h} 06^{\rm m} 03.6^{\rm s},
\delta_{2000} = -25^{o} 45' 46"$ containing the QSO Q 0103-260
(z = 3.36) was chosen as the FDF. The DSS print in Figure 1 shows a 
comparison of the FDF and the HDF-S.

\begin{figure}
\centerline{\hbox{
\psfig{figure=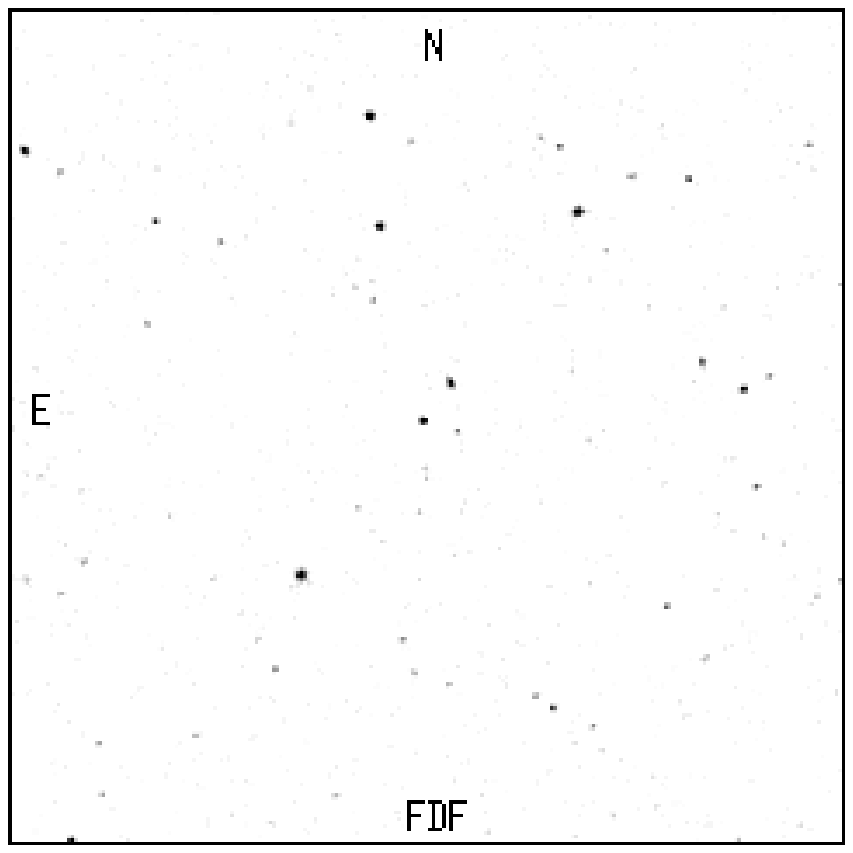,width=5.7cm,height=5.5cm,clip=t}
\hspace*{.2cm}
\psfig{figure=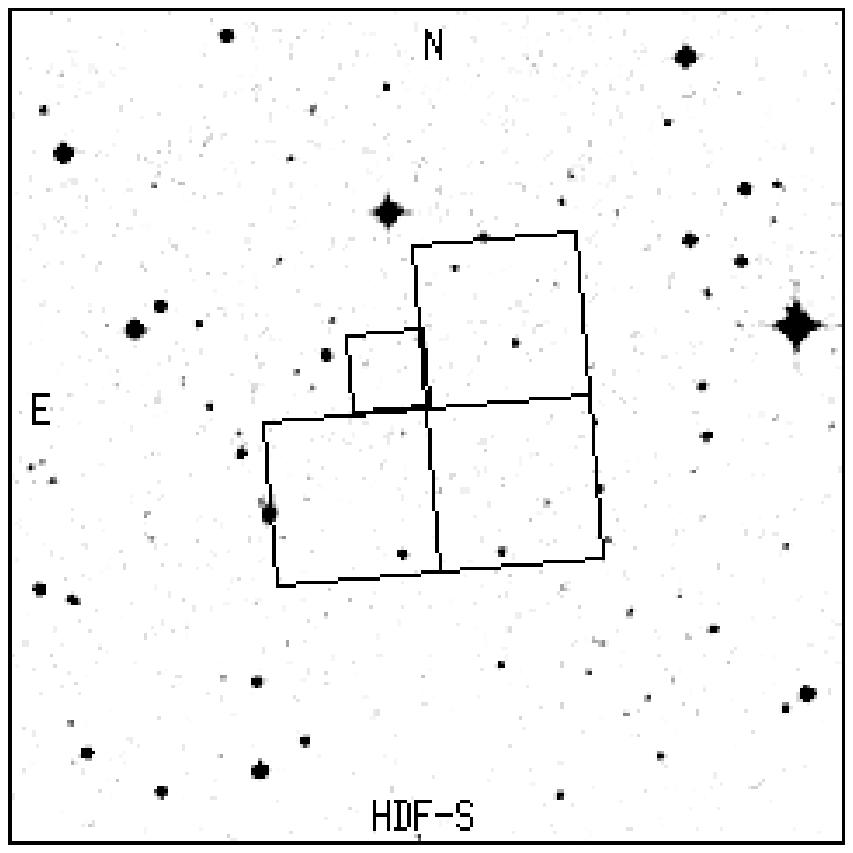,width=5.7cm,height=5.5cm,clip=t}
}}
\caption [] {DSS plots of the FDF and of a field of the same size surrounding
the HDF-S. Note the much lower surface density of bright foreground 
objects and the absence of bright stars in the FDF region.}
\end{figure}

\section{Projects within the FDF}

The main aim of the FDF is to study the galaxy evolution in the 
redshift range between z = 1 - 5 (morphology, chemical abundances,
star formation rates, number counts etc..). More specific projects address the
evolution of the Tully-Fisher relation in field spirals between z = 0.2 - 1 
and the evolution of field ellipticals between z = 0.2 - 0.6.
Additionally, the QSO host and its environment (cluster properties
and ${\rm L}_{\alpha}$-galaxies) as well as absorption line systems along the 
line of sight to the QSO will be studied. Finally, galaxy-galaxy lensing
signals and QSO statistics will be investigated.

\section{Observations}

The imaging of the FDF was carried out mainly with FORS1 during 
5 runs between August and December 1999. During those runs, 
observations in UBgRIz and two narrow band filters (530\_25 and
485\_37, the former centered at Ly$\alpha$ at the QSO's redshift, to 
search for Ly$\alpha$-galaxies around z = 3.36) were taken. During periods of 
unfavourable seeing (which unfortunately occurred several times) 
multi-object spectroscopy of $\sim$ 120 galaxy candidates in the FDF was made.
The galaxy candidates were selected from preliminary photometric redshifts
based on optical data of the FDF taken during the field selection procedure.

Images of the FDF in J and Ks were acquired with the NIR camera SOFI at the NTT
in October 1999. The main purpose of the NIR images was to
optimize the photometric redshifts for follow-up spectroscopy. 
The majority of the multi-object spectroscopic observations of
high-redshift galaxy candidates was carried in two runs within three 
weeks after the annual meeting of the German Astronomical Society
2000 in Bremen. During these runs excellent 
spectra of $\sim$ 230 galaxy candidates with photometric redshifts 
mainly between z = 1 - 5 (and a few z = 6 candidates) could be collected.

\section{Results}

\subsection{Imaging}

An overview about the imaging results for each individual filter (as of
September 2000) along with the total integration time, the PSF FWHM of the
summed image, the field of view (FOV) and 5$\sigma$ completeness limits for
galaxies are presented in Table 1. The FOVs of the individual optical filters 
differ somewhat, which is caused by few positioning errors of the telescope
and/or the fact that the images have been taken in different runs.
While the total integration times are within our expectations,
the completeness limits are not. This mainly due a relatively  
low efficiency of 
the telescope and the CCD (resulting in a loss of approx. 0.4mag) at the time
of the observations. 
Moreover, the seeing was on average more worse than expected.
Still, we detected $\sim 10^4$ galaxies in the FDF. For example, 
we detected $\sim$ 6500 galaxies in B band($\sim$ 4100 up to B = 27.75), 
$\sim$ 8600 galaxies in R band ($\sim$ 5100 up to R = 26.75) and
$\sim$ 7500 galaxies in I band ($\sim$ 4600 galaxies up to I = 26.25).
The total integration time in J and Ks was relatively low, which is due to the
relatively 

\begin{table}[h]
\caption[]{Overview about the imaging observations.}
\begin{center}
\begin{tabular}{c|rccc}
\hline
 & & & & \\
Filter & Time     & FWHM & FOV & Completeness (5 $\sigma$)\\ 
       &  [hours] & [''] & ['] & [mag]\\ 
 & & & & \\
\hline
 & & & & \\
U       & 10.7 & 1.15 & 6.2 $\times$ 6.4 & 25.50\\
B       & 4.4  & 0.95 & 5.8 $\times$ 5.8 & 27.50\\
g       & 5.0  & 1.00 & 6.3 $\times$ 6.4 & 27.00\\
R       & 5.8  & 0.85 & 6.4 $\times$ 6.4 & 26.75\\
I       & 4.0  & 0.60 & 6.1 $\times$ 5.8 & 26.25\\
z       & 3.5  & 0.75 & 6.8 $\times$ 6.8 &     tbd    \\
530\_25 & 4.0  & 0.80 & 6.8 $\times$ 6.8 &     tbd    \\
485\_37 & 3.0  & 0.85 & 6.8 $\times$ 6.8 &     tbd     \\
 & & & & \\
\hline
 & & & & \\
J       & 1.7  & 0.80 & 6.8 $\times$ 7.9 & 22.50\\
K       & 1.7  & 0.75 & 6.8 $\times$ 7.9 & 20.50\\
 & & & & \\
\hline
\end{tabular}
\end{center}
\end{table}

\noindent small FOV of SOFI as compared to FORS. Therefore a mosaic of 
four pointings was necessary. Since the SOFI subfields had some overlap, 
some parts of the FDF (including the QSO)
were exposed much longer (up to 6.8h).

\begin{figure}
\centerline{\hbox{
\psfig{figure=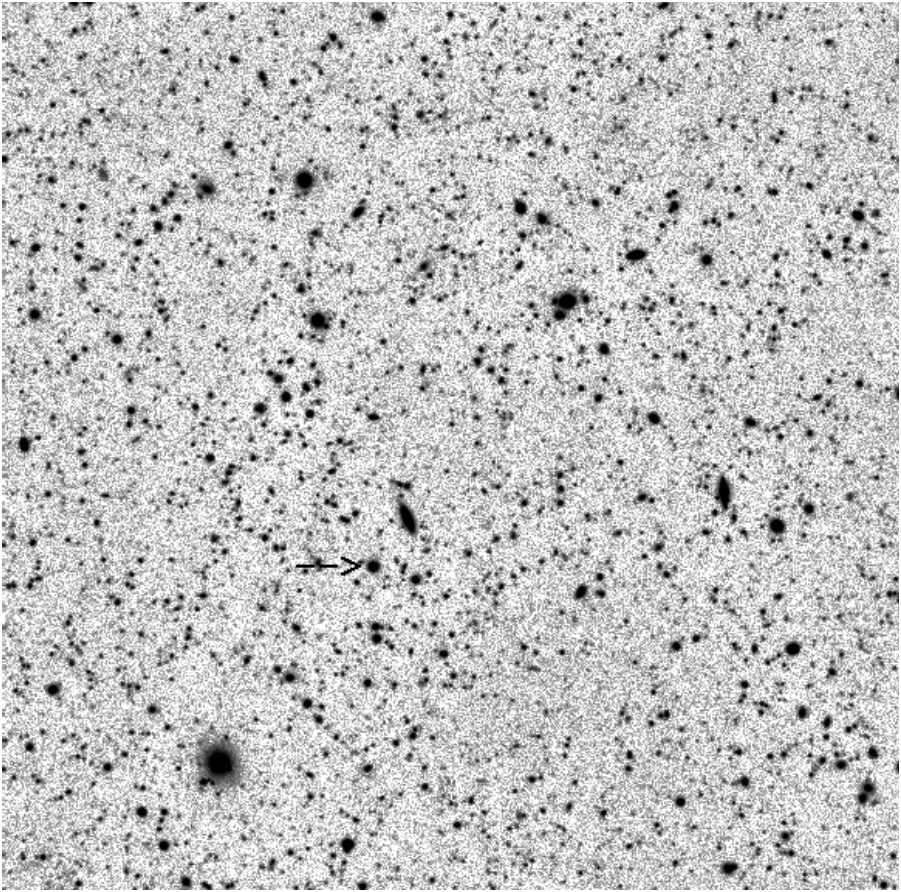,height=11.8cm,clip=t}}}
\caption [] {The FDF in B band from FORS observations. 
This frame contains $\sim$ 6500 galaxies.
Integration time is 4.4h, FWHM = 0.95", FOV = 5.8' $\times$ 5.8'. 
North is up, east to the left. The QSO 
is marked with an arrow.}
\end{figure}

In Figure 2, the FDF in B band is shown. Obviously, the galaxies 
are homogeneously distributed in the FDF. The brightest object is an 
elliptical galaxy in the lower southeastern part of the FDF with B $\sim$ 18.5.
Unfortunately, there is still a (fortunately not very rich) galaxy 
cluster at z $\sim$ 0.3 in the southwestern corner of the FDF. 

\begin{figure}
\centerline{\hbox{
\psfig{figure=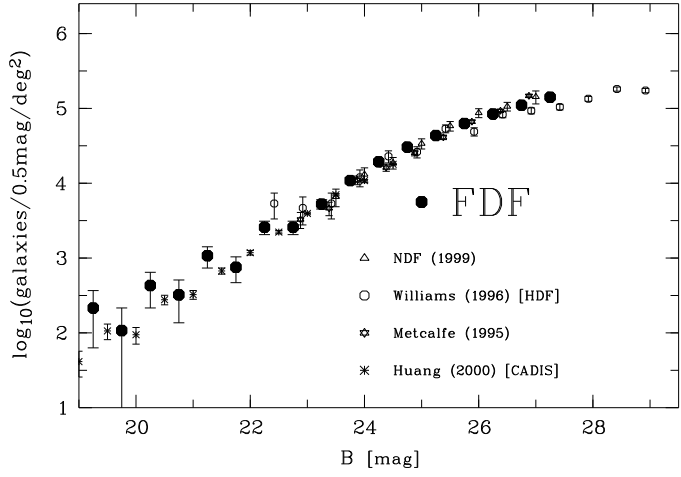,width=10cm,angle=-90,clip=t}}}
\caption [] {Galaxy number counts of the FDF in B band as compared
to other deep surveys.}
\end{figure}

\begin{table}
\caption[]{Limiting magnitudes of the FDF compared to those of the 
NTT Deep Field and the HDF-N.}
\begin{center}
\begin{tabular}{c|ccc|ccc|cc}
\hline
 & & & & & & & & \\
 & \multicolumn{3}{c|}{FDF} & \multicolumn{3}{c|}{NTT DF} 
& \multicolumn{2}{c}{HDF-N} \\ 
& & & & & & & & \\
\hline
& & & & & & & & \\
 & lim         & FWHM & Time   & lim & FWHM & Time & lim & Time \\ 
       & 5 $\sigma$ & ['']  & [hours]& 5 $\sigma$ & [''] & [hours] 
& 10 $\sigma$ & [hours] \\
 & & & & & & & &\\
\hline
 & & & & & & & &\\
U       & 25.50 &1.15 & 10.7 &      &&& 27.0 &  43\\
B       & 27.50 &0.95 & 4.4  & 26.9 &0.90&14.7& 27.9 & 33.5 \\
g       & 27.00 &1.00 & 5.0  & 26.5 &0.83&6.5& 28.0 &  30\\
R       & 26.75 &0.85 & 5.8  & 25.9 &0.83&6.5&      &  \\
I       & 26.25 &0.60 & 4.0  & 25.3 &0.70&4.5& 27.6 &  34\\
 & & & & & & & &\\
\hline
\end{tabular}
\end{center}
\end{table}

The "quality" of the FDF can also be demonstrated by the 
galaxy number counts. As an example, the galaxy number counts in B band 
for the FDF are compared to those derived by
Metcalfe et al. 1995 (WHT/INT observations), 
Williams et al. 1996 (HDF-N), Arnouts et al. 1999 (NTT Deep Field)
and Huang et al. 2000 (CADIS). In general there is an excellent agreement
over almost 9 mag, with some scatter at the bright end. This might be due to 
the contamination of the galaxy cluster in the southeastern corner of the FDF,
which is still small.

It is also tempting to compare the completeness limits of the FDF to those of 
other Deep Fields. These are shown for the  broad-band optical filters 
in Table 2 in comparison to the HDF-N and the NTT Deep Field.
Whereas the FDF is $\sim$ 0.5-1 mag deeper as the NTT Deep Field,
it is 0.4 - 1.3 mag less deep as the HDF-N. Given the 
loss of 0.4 mag due to the low efficiency of the VLT and the CCD, the FDF 
is indeed comparable to the HDF-N except in the I filter. 
This is no surprise, however,
since atmospheric effects (airglow) does not effect the space-borne HST.
It should be noted that the FOV of the FDF is a factor of 6 larger 
($\sim$ 36 sq. arcmin.) as compared to the HDF-N and the NTT Deep Field
($\sim$ 5.6 sq. arcmin.). Finally, it should be emphasized that the
integration times are comparable to 
the NTT Deep Field, whereas the HDF-N was exposed between
4 and 8 times longer.

\subsection{Photometric redshifts}

Photometric redshifts were estimated 
by fitting template spectra of different galaxy types/stars and varying redshift
to the measured fluxes on the optical and NIR images of the objects
(see Bender 2001 for details). In order to minimize biases 
(e.g.by absorption through dust) an I band selected sample was chosen.
Photometric redshifts and object types could be estimated 
for $\sim$ 3800 objects. Most objects of the I-band selected sample
are Im/Irr-type galaxies, about 15\% are early (E/S0) or late-type
(Sa/Sb/Sc) galaxies, about 5\% (190) are most likely stars.
The major fraction of the galaxies have photometric
redshifts between z = 0 and 3, a few objects have photometric redshifts
between z = 5 and 6 (some, perhaps all of which might be late-type
M or L stars). In general, the early-type galaxies haver lower redshifts than
the late-type galaxies, which have a similar redshift distribution 
as the Im/Irr-galaxies. This is illustrated in
Figure 4, where the distribution of the full galaxy sample and two
subsamples are shown. For comparison, fits to the photometric redshift 
distribution of the HDF-S and HDF-N are shown as well. In general, 
the distributions of the photometric redshifts between the FDF and the 
two HDFs are quite similar.
The reliability of the photometric redshifts can be tested best
by a comparison to spectroscopic redshifts from real data.
This was possible, since spectra of $\sim$ 120 galaxy candidates could be 
collected already during the imaging part of the FDF (see above). 
A comparison is presented in Figure 5. 
As can be seen, there is a very good consistency
between the photometric and spectroscopic redshifts (the former having
errors of typically $\leq$ 10\%), with only a few outliers. 
Hence, the photometric redshifts are 
an excellent preselection criteria
for follow-up spectroscopy the I-band selected galaxy sample.   

\begin{center}
\begin{figure}
\centerline{\hbox{
\psfig{figure=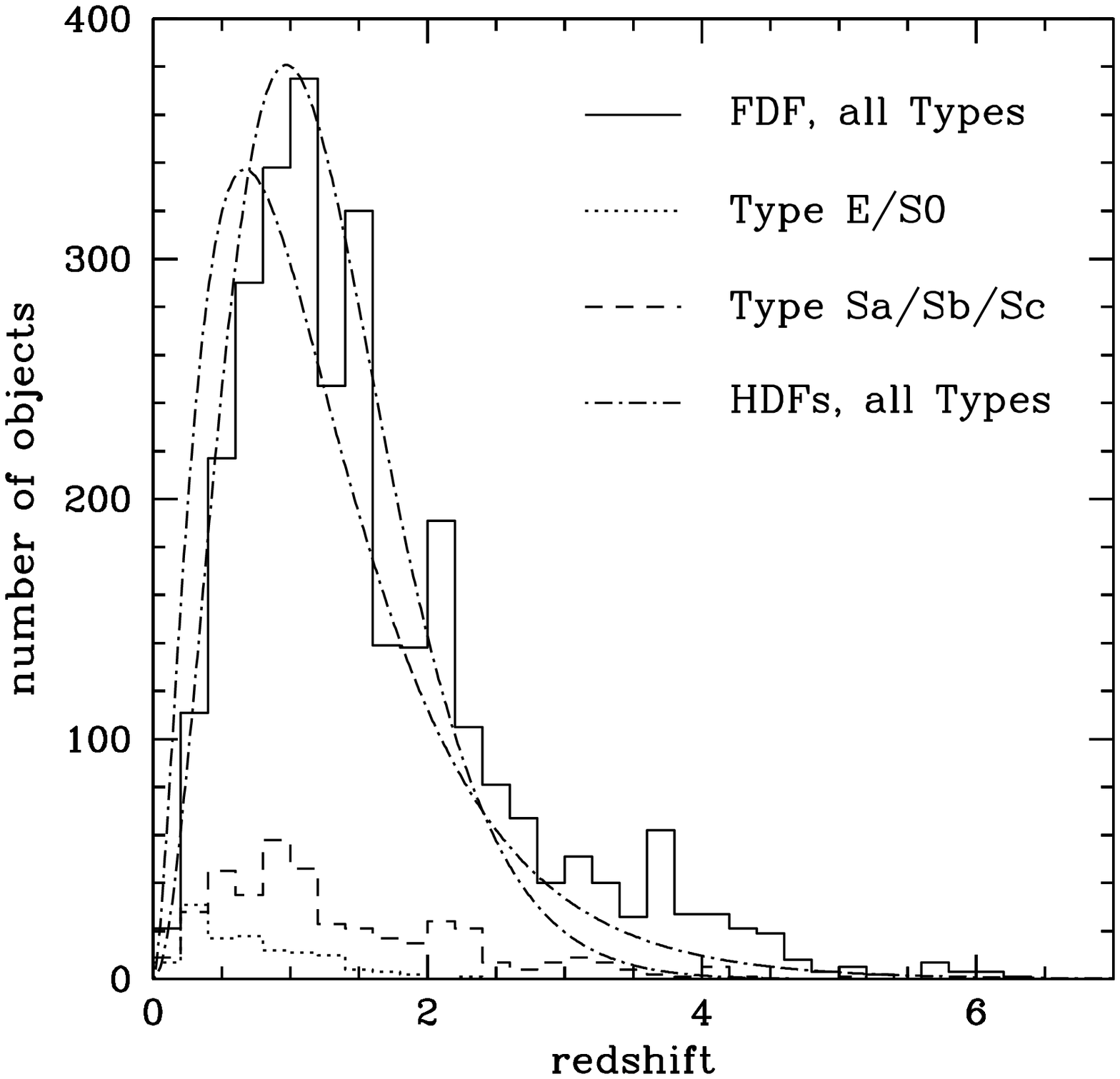,height=8cm,clip=t}}}
\vspace*{-.5cm}
\caption [] {Distribution of the photometric redshifts estimated from an
I-band selected sample (3600 galaxies) in the FDF. The dotted line represents
the early-type galaxies, the dashed line the late-type  
galaxies  and the solid line the sum of all galaxy types
(early-type, late-type and Im/Irr). For comparison fits to the distributions
of photometric redshifts from the HDFs are shown.}

\centerline{\hbox{
\psfig{figure=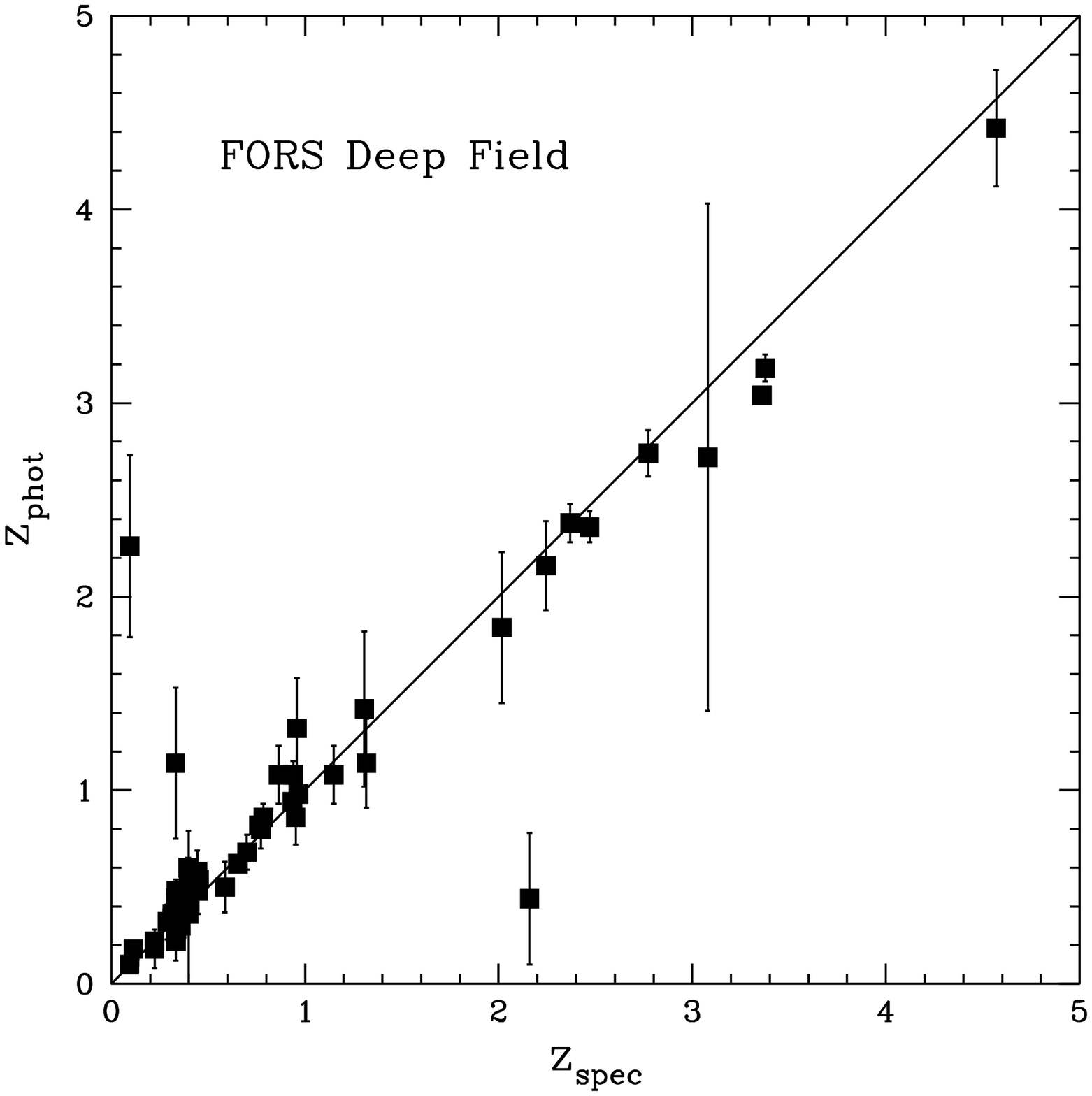,height=8cm,clip=t}}}
\vspace*{-.5cm}
\caption [] {Comparison of spectroscopic and photometric redshifts for
$\sim$ 100 galaxies in the FDF. Note the very good agreement for almost all
galaxies.}
\end{figure}
\end{center}

\subsection{Spectroscopy}

During the imaging part of the FDF spectra of $\sim$ 120 (relatively bright) 
galaxy candidates could be collected. Most of the objects (65) turned out to
be galaxies at low redshift (z $<$ 1), stars (20) or unidentified
objects (20). Nevertheless, also 15 galaxies with z = 1 - 4 were found,
thus illustrating the power of the VLT even under relatively poor
observing conditions. The spectrum of a z = 2.773 galaxy from those 
observations is shown in Figure 6. This galaxy is dominated by a strong
(rest frame) UV continuum, has no strong emission lines, but many
high ionisation metallic lines in absorption indicating 
strong starburst activity.
For illustration we display the appearance of this galaxy in B, R, I and K
on the same figure. On the B-band frame the galaxy is clearly visible with some
hint of an elongation towards the south, which is more prominent 
in the R-band image. A nearby galaxy (early-type, photometric 
redshift $\sim$ 0.36) to the southeast can also be seen. 
On the I-band image, the southern
extension is somewhat separated from the z = 2.773 galaxy. Moreover,
due to the good seeing of that image, an extension towards the northeast 
become visible. Although the K-band image is not very deep,
the three components (z = 2.773 galaxy, companion at 
${\rm z}_{\rm phot}$ = 0.36 and southern extension) are present
as well. Although this is a bit speculative,  
the starbursting z = 2.773 galaxy might be in interaction with either 
extension to the south or north or in some merger stage. 
This has frequently been observed in galaxies at high redshift.

\begin{figure}
\centerline{\hbox{
\psfig{figure=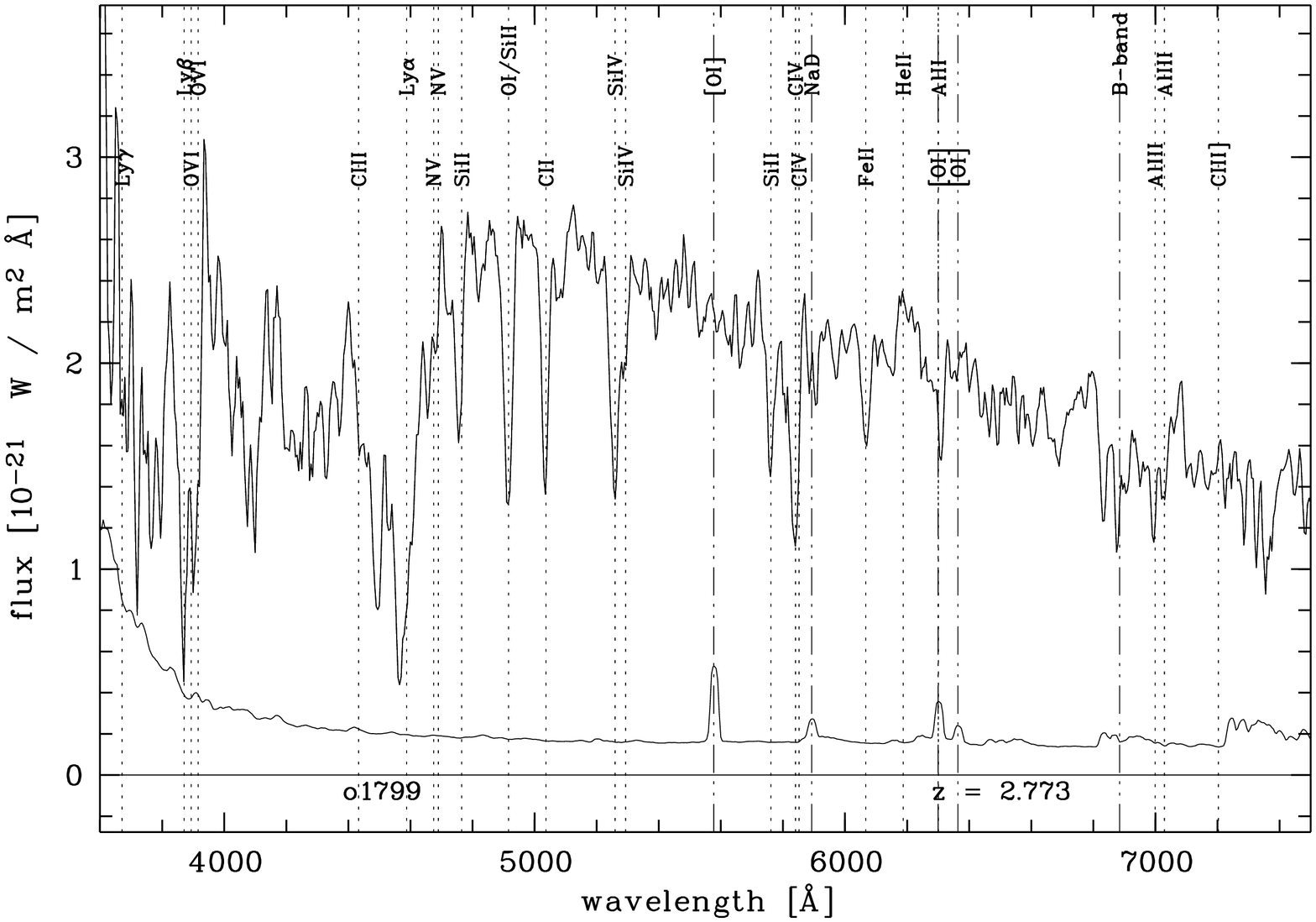,width=10cm,clip=t,angle=-90}}}
\vspace*{.5cm}
\centerline{\hbox{
\psfig{figure=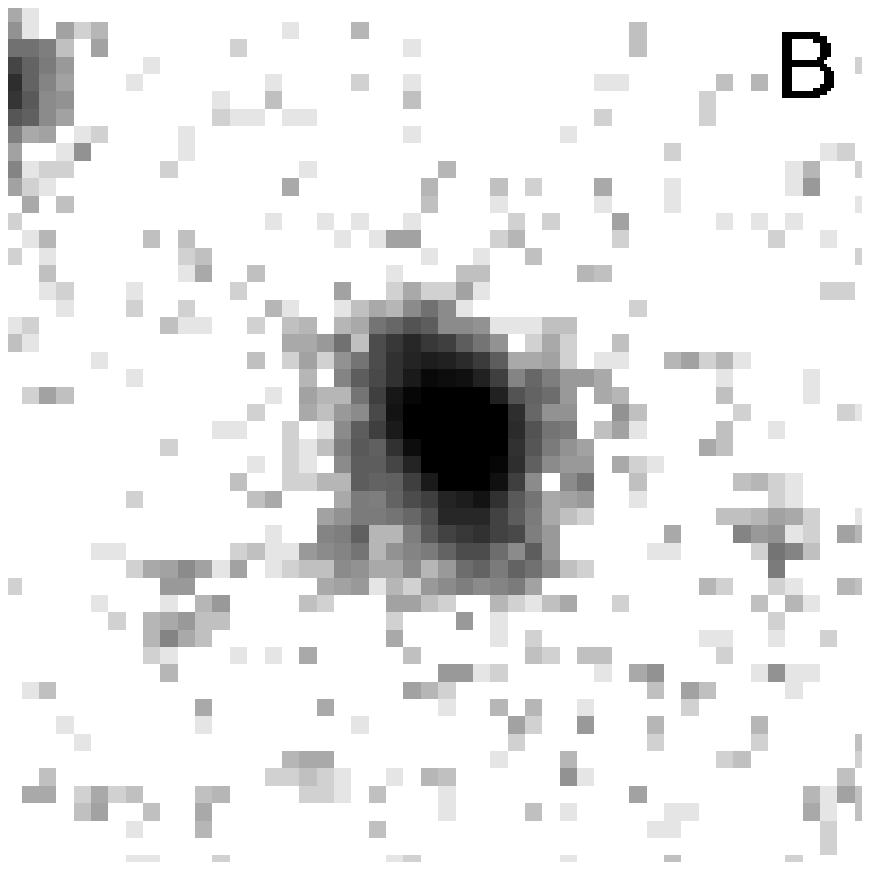,width=4cm,clip=t}
\hspace*{.2cm}
\psfig{figure=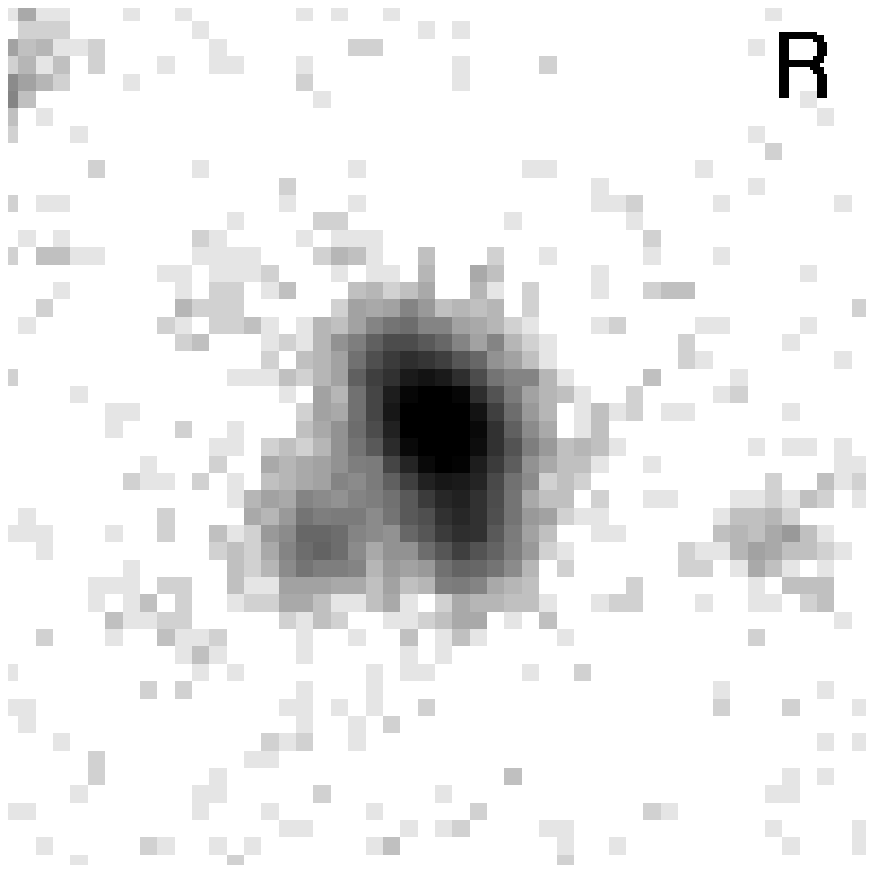,width=4cm,clip=t}
}}
\vspace*{.2cm}
\centerline{\hbox{
\psfig{figure=heidtfig6d.ps,width=4cm,clip=t}
\hspace*{.2cm}
\psfig{figure=heidtfig6e.ps,width=4cm,clip=t}
}}

\caption [] {Top) Spectrum of a z = 2.773 FDF galaxy. Note the strong
UV (rest frame) continuum and the absence of strong emission lines.
Bottom) Images of the z = 2.773 galaxy in B, R, I and Ks.
North is up, east to the left. FOV is 10" $\times$ 10".
See text for details.}
\end{figure}

\begin{figure}
\centerline{\hbox{
\psfig{figure=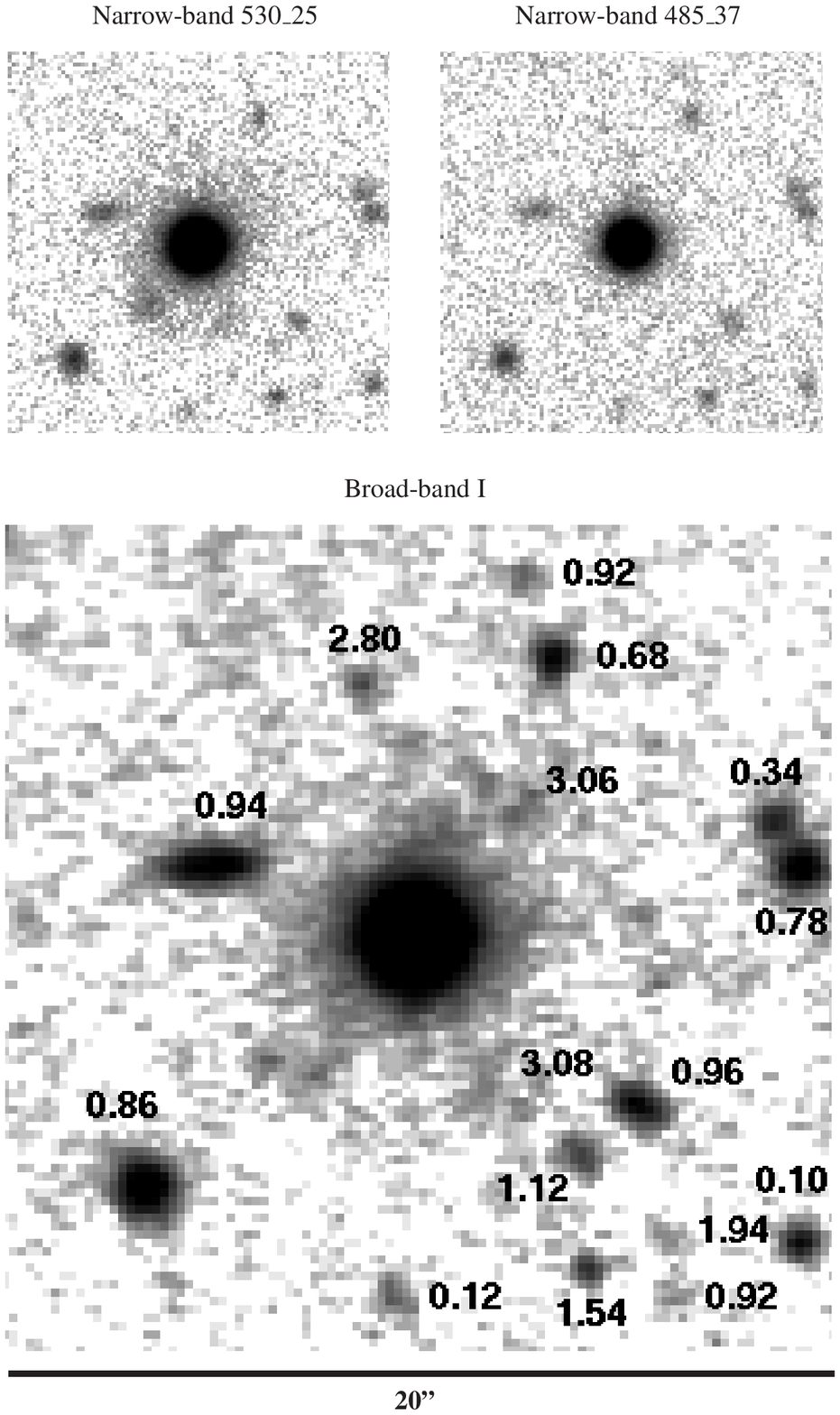,width=9.5cm,clip=t}}}
\caption [] {Narrow-band image of the QSO environment centered at the
redshifted Ly$\alpha$ of the QSO (530\_25, top left) compared to the second 
narrow-band image (485\_37, top right) and the broad-band I image (bottom).
Note the two objects a few arcsec southeast and southwest of the QSO on the
530\_25 image, which are not visible on the 485\_37 image. A further object
at photometric z = 3.06 is present on the broad-band I image. These three
sources may form a small group with the QSO or are along the line-of-sight to
the QSO.}
\end{figure}

\subsection{Environment of QSO Q0103-260 (z = 3.36)}

While broad-band imaging of the FDF 
was mainly carried out for determining photometric redshifts 
(target selection for spectroscopic follow-up) and morphological studies, 
narrow-band imaging was conducted specifically to search
for Ly$\alpha$-galaxies at the QSOs redshift. The first results from the 
latter are encouraging. In Figure 7 the environment of the QSO 
in the narrow-band image centered at the redshifted Ly$\alpha$ of the QSO 
(530\_25) are compared to the image taken with another narrow-band filter
(485\_37) and the broad-band I image. On the broad-band I image the 
photometric redshifts of individual galaxies are displayed.

Several interesting features can bee seen on that Figure. 
On the 530\_25 filter image two objects a few arcsec 
southeast and southwest of the QSO are detected, which are not visible 
on the 485\_37 filter image, but faintly present on the
broad-band I image. Moreover, one of the two objects has a photometric
redshift (z = 3.08), close to the QSOs redshift. Additionally, 
an object with a similar photometric redshift (z = 3.06) was found,
which is  a few arcsec to the northwest of the QSO. 
Given an error of $\leq$ 10\% for
photometric redshifts of such faint objects (mainly due to the errors
of the photometry) the QSO might well be surrounded by a group of 
at least 3 galaxies,
two of which having relatively strong Ly$\alpha$-emission. Alternatively,
the three galaxies (or at least some of them) might be along the line of sight
to the QSO, thus being responsible for absorption-line systems in the
spectrum of the QSO. Indeed, some absorption-line systems close to z = 3.36
have been detected in a low-resolution spectrum of the QSO (Figure 8). 

\begin{figure}
\centerline{\hbox{
\psfig{figure=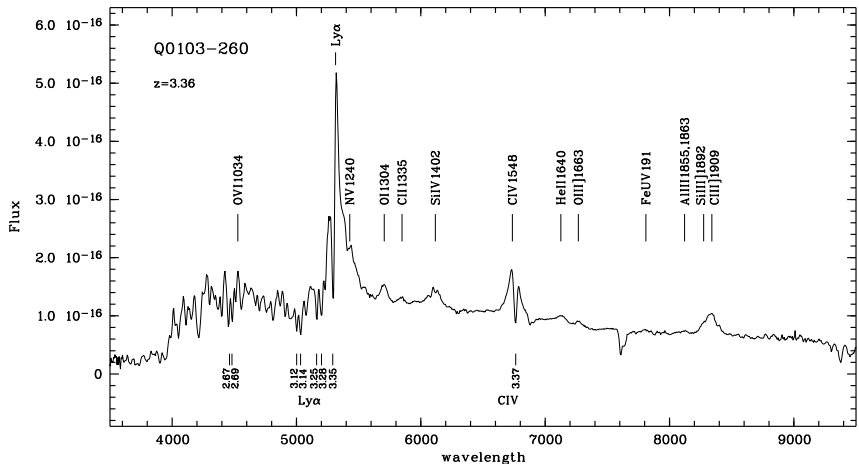,width=11.5cm,angle=-90,clip=t}}}
\caption [] {Low-resolution spectrum of QSO Q0103-260 at z = 3.36.
Several prominent emission lines typical for QSOs as well as  
a few absorption absorption lines and their corresponding redshifts 
are marked.}
\end{figure}

Spectroscopy of these three sources with FORS at the VLT will clarify
the possible physical association with the QSO. Since the sources are
relatively faint (${\rm m}_{\rm I} = 25$) this is a challenging task even for
an 8m class telescope. Finally, a high-resolution  spectrum of the QSO with 
UVES at the VLT will allow the identification of the absorption-line 
systems more accurately. Joined together, the nature and physics of these
objects can then be studied in detail.

\section{Summary and outlook}

The first observations of the FDF started a bit more than one year ago.
During that time, the imaging part was carried out. The analysis of the data
(photometric redshifts, galaxy number counts etc.) is essentially completed. 
Although the observing conditions (seeing) were not as good as anticipated 
and although some light was lost due to the lower efficiency of the telescope
(degrading CCD etc.) the FDF is one of the deepest view of a region on the 
sky ever made from ground, nearly comparable as the HDFs, but with a factor of
6 larger area surveyed. Spectra of 
approximately 120 objects in the FDF were collected as well, which 
were primarily used to prove the reliability and to optimize 
the photometric redshift code. Some of these spectra are also useful for 
the main aim of the FDF - the study of the evolution of galaxies at early 
epochs. The first analysis of the narrow-band images is encouraging. 
Three objects, presumably galaxies, nearby the QSO could be identified, 
which are either physically associated with the QSO, or are along the 
line-of-sight to the QSO thus responsible for the absorption line systems in
the QSO spectrum.

Since the annual meeting of the German Astronomical Society in autumn 2000 in
Bremen, the major part of the spectroscopical observations was carried out. 
Spectra for more than 200 high-z galaxy candidates could be taken with 
FORS at the VLT. The preliminary results confirm the reliability of the
photometric redshift code further. This will allow to study  
the properties of the high-z galaxies statistically, which are unobservable 
spectroscopically due to their faintness even with the VLT. 
On the other hand, the high-quality spectra for more than
200 galaxies in the redshift range z = 1 - 5 (with the current highest-z
galaxy at a redshift of z $\sim$ 4.6 (S. Noll, priv. com.)) form an excellent 
database for future investigations of the evolution of galaxies 
in the early universe. 

\section{Acknowledgments}

It is a pleasure to thank the organizers of the annual meeting of the
German Astronomical Society 2000 in Bremen for the opportunity to present 
the FORS Deep Field project. The generous support by the Paranal and 
La Silla staff 
during the various observing runs is gratefully acknowledged. This work was
supported by the Deutsche Forschungsgemeinschaft (SFB 375, SFB 439),
the VW foundation and the German Federal Ministry of Science and Technology
with ID-Nos. 05 2HD50A, 05 2GO20A and 05 2MU104.

\vspace{0.7cm}
\noindent
{\large{\bf References}}
{\small

\bref Appenzeller I., Rupprecht G., et al., 1992, "FORS, the Focal 
Reducer for the VLT", The Messenger 67, p. 18 

\bref Arnouts S., D'Odorico S., Cristiani S., et al., 1999, A\&A 341, 641

\bref Bender R., 2001, "The FORS Deep Field: Photometric redshifts and object
classification", Workshop on "Deep Fields", in press

\bref Huang J.S., Thompson D.J., K\"ummel M.W., et al., 2000, 
A\&A submitted 

\bref Kauffmann G., 1996, MNRAS 281, 487

\bref Koo D., 2000, "Exploring distant galaxy evolution: Highlights with
Keck", Reviews in Modern Astronomy 13 (R.E. Schielicke, ed.), p. 173, 

\bref Larson R.B., 1974, MNRAS 166, 585

\bref McCracken H.J., Metcalfe N., Shanks T., et al., 2000, MNRAS 311, 707

\bref Meisenheimer K., Beckwith S., Fockenbrock R., et al., 1998, 
"The Calar Alto Deep Imaging Survey (CADIS)", in "The young universe: 
galaxy formation and evolution at intermediate and high redshift", 
S. O'Dorico, A. Fontana, E. Giallongo (eds.), ASP Conf. Ser. 136, p. 134

\bref Metcalfe N., Shanks T., Fong R., Roche N., 1995, MNRAS 273, 257

\bref V\'{e}ron-Cetty M.P., V\'{e}ron P., 1997, A catalogue of Quasars and
Active Nuclei (7. edition)

\bref Williams R.E., Blacker B., Dickinson M., et al., 1996, 
AJ 112(4), 1335
}

\vfill

\end{document}